\documentclass[final,5p,times,twocolumn,nopreprintline]{elsarticle}
\usepackage{amsmath,slashed,booktabs,dsfont}
\usepackage{graphicx,graphics}
\usepackage{dcolumn}
\usepackage[hyperfootnotes=false]{hyperref}
\usepackage{xspace}
\usepackage{color}
\usepackage{balance}
\usepackage{multirow}
\usepackage{multicol}

\usepackage{fancyhdr}
\fancypagestyle{firstpage}{%
	
	\lhead{}
	\rhead{}
}
\newcommand{\GeV}{\,\text{GeV}}

\newcommand{\Order}{\mathcal{O}}

\renewcommand{\Im}{\text{Im}\,}
\renewcommand{\Re}{\text{Re}\,}

\allowdisplaybreaks[1]

\begin{document}

\renewcommand{\theequation}{\arabic{equation}}

\begin{frontmatter}
 
\title{Four-fermion operators, $Z$-boson exchange, and $\tau$ lepton dipole moments}

\author[Bern]{Jo\"el Gogniat}
\author[Bern]{Martin Hoferichter}
\author[Bern]{Gabriele Levati}

\address[Bern]{Albert Einstein Center for Fundamental Physics, Institute for Theoretical Physics, University of Bern, Sidlerstrasse 5, 3012 Bern, Switzerland}

\begin{abstract}
Asymmetry measurements in $e^+e^-\to\tau^+\tau^-$ constitute a promising avenue to obtain competitive constraints on the $\tau$ dipole moments, the anomalous magnetic moment $a_\tau$ and the electric dipole moment $d_\tau$, especially, once a polarized electron beam becomes available, as possible at a future polarization upgrade of the SuperKEKB collider. While the main challenges concern the measurement of these asymmetries and the calculation of radiative corrections at the relevant level of precision, at subleading orders also electroweak effects and the potential impact of four-fermion operators parameterizing other beyond-the-Standard-Model scenarios besides those described by dipole operators need to be taken into consideration. Here, we show that $Z$-boson contributions arise at the level of $\simeq 3\times 10^{-6}$, while we estimate the largest possible effect from four-fermion operators as $\simeq 10^{-5} C \, v^2/\Lambda^2$.
In addition, we observe that four-fermion-operator insertions at the loop level can probe Wilson coefficients that are otherwise not  constrained directly, and that the imaginary part generated by insertions of the dipole operator at loop level opens another potential avenue towards a determination of $a_\tau$ without the need for a polarized electron beam. Despite the inherent loop suppression, a measurement of the required normal asymmetry $A_N^\pm$ with a precision of $\lesssim 10^{-5}$ would allow one to probe the Schwinger term, which could define an intermediate goal to be realized in the current setting at Belle II.
\end{abstract}

\end{frontmatter}

\thispagestyle{firstpage}

\section{Introduction}

A promising avenue to probe the anomalous magnetic and electric dipole moments of the $\tau$ lepton, $a_\tau$ and $d_\tau$, relies on the measurement of suitably defined asymmetries in the process $e^+ e^- \to \tau^+ \tau^-$ at $B$ factories, as originally proposed in Refs.~\cite{Bernabeu:2004ww,Bernabeu:2006wf,Bernabeu:2007rr,Bernabeu:2008ii} to be realized at the lower $\Upsilon$ resonances. Recently, this proposal has been pursued in the context of the physics program of a future polarization upgrade of the SuperKEKB collider~\cite{Crivellin:2021spu,USBelleIIGroup:2022qro,Aihara:2024zds}, as the asymmetries that become accessible with a polarized electron beam would allow one to test $a_\tau$ at the level of $10^{-5}$ and possibly further, reaching far beyond current limits~\cite{CMS:2024qjo,ATLAS:2025oiy} that still fall short of the QED Schwinger term~\cite{Schwinger:1948iu}. Likewise, access to electron polarization would imply improved sensitivity to $d_\tau$~\cite{Belle:2021ybo}.
To make this program work down to a precision of $\mathcal{O}(10^{-6})$ in $a_\tau$, radiative corrections need to be controlled at two-loop order~\cite{Crivellin:2021spu},
and Monte-Carlo tools developed at the same level of accuracy~\cite{Gogniat:2025eom,Banerjee:2020rww,Ulrich:2025fij}. In Ref.~\cite{Gogniat:2025eom}, the complete one-loop analysis in QED is provided, and the extension to the next order is ongoing.

The Standard-Model (SM) prediction for $a_\tau$ is known at $\simeq 4\times 10^{-8}$~\cite{Eidelman:2007sb,Eidelman:2016aih,DiLuzio:2024sps,Hoferichter:2025fea}, so that there is ample room before theory uncertainties in $a_\tau$ itself start limit the sensitivity to physics beyond the SM (BSM). Realistic scenarios, even in the case of chiral enhancement~\cite{Giudice:2012ms,Crivellin:2021rbq,Athron:2025ets}, strongly suggest that a target precision $\lesssim 10^{-5}$ is necessary to obtain meaningful bounds~\cite{Crivellin:2021spu} and render $a_\tau$ a viable precision probe that can help elucidate the chirality and flavor structure of potential BSM contributions to lepton anomalous magnetic moments~\cite{Aoyama:2020ynm,Fan:2022eto,Muong-2:2025xyk,Aliberti:2025beg,Hertzog:2025ssc}.  Similarly, $d_\tau$ is the least explored one among the lepton electric dipole moments, and in view of the ongoing efforts for $d_\mu$~\cite{Abe:2019thb,Aiba:2021bxe,Adelmann:2025nev,Crivellin:2018qmi}, as well as the stringent limits for $d_e$~\cite{Roussy:2022cmp,ACME:2018yjb},  the sensitivity gap would be bound to increase in the near future without concurrent improvements for $d_\tau$.   

A common feature of several strategies to constrain $a_\tau$ and $d_\tau$, be it in $e^+e^-\to\tau^+\tau^-$~\cite{DELPHI:2003nah,Gonzalez-Sprinberg:2000lzf}, radiative $\tau$ decays~\cite{Eidelman:2016aih}, LHC processes~\cite{Galon:2016ngp,Koksal:2017nmy,Gutierrez-Rodriguez:2019umw,Beresford:2019gww,Dyndal:2020yen,ATLAS:2022ryk,CMS:2022arf,Haisch:2023upo,Shao:2023bga,Beresford:2024dsc,Dittmaier:2025ikh,CMS:2024qjo,ATLAS:2025oiy}, or future high-energy colliders~\cite{Buttazzo:2026amk}, is that the dipole moments are not measured directly, but instead modifications of the effective $\tau\tau\gamma$ vertex are being probed.
Accordingly, the actual observable is the dipole form factor at the center-of-mass energy $\sqrt{s}$ of the machine, and an effective-field-theory (EFT) argument is needed to extract constraints on $a_\tau$ and $d_\tau$ themselves, which applies as long as the BSM scale $\Lambda$ is much larger than $\sqrt{s}$. In the case of asymmetries in $e^+e^-\to\tau^+\tau^-$, the difference between their measurements and SM predictions can be directly interpreted in terms of $a_\tau^\text{BSM}$ and $d_\tau^\text{BSM}$ (in which case the SM contribution is negligibly small~\cite{Yamaguchi:2020eub}). For light BSM degrees of freedom, the same measurements still permit an interpretation in terms of $a_\tau$ and $d_\tau$, but the connection becomes model dependent, see Refs.~\cite{Hoferichter:2025ijh,Hoferichter:2025zjp} for a comprehensive analysis. In particular, it was found that, apart from regions of accidental cancellations, the sensitivity remains roughly at the same level as set by the asymmetry measurements. Moreover, it was observed that the imaginary part generated for $s>4m_\tau^2$ entails comparable sensitivity to $a_\tau$ and $d_\tau$ for mediator masses around $\sqrt{s}$, before the EFT decoupling sets in. This alternative strategy can be realized already in the current setting at Belle II~\cite{Hoferichter:2025ijh}, since the imaginary parts can be determined from the normal asymmetry without the need  for electron polarization.

Further subtleties in the interpretation that could become relevant when approaching the level of $10^{-6}$ and beyond concern possible contamination from electroweak effects and other effective operators in the low-energy EFT (LEFT), both of which we address in this Letter. 
After reviewing the basic formalism in Sec.~\ref{sec:formalism}, we first discuss the exchange of a $Z$ boson in Sec.~\ref{sec:Zboson}. 
A detailed study of electroweak corrections 
 to $e^+e^-\to\tau^+\tau^-$ for the unpolarized case was already performed in Ref.~\cite{Kollatzsch:2022bqa}, while partial results for the asymmetries were provided in
Refs.~\cite{Bernabeu:1994wh,Bernabeu:2006wf,Bernabeu:2007rr}; here, we work out the complete $Z$-boson contributions to the asymmetries of interest and quantify the size of the corrections for Belle II energies.    
Next, while the asymmetries are constructed in such a way as to isolate modifications to dipole operators, also other LEFT effective operators  could contribute, and one needs to ensure that BSM contributions besides those of dipole type do not impact the asymmetries at the relevant level. Here, we provide such an analysis, see Sec.~\ref{sec:4F_tree}  for four-fermion operators at tree level, generating effects both
 by SM--BSM interference and exclusively by heavy BSM contributions.
In addition, in Sec.~\ref{sec:4F_loop} we discuss one-loop effects involving four-fermion operators. As an application along the same lines, we observe that loop-level insertions of dipole operators generate an imaginary part as well, which again can be inferred from the normal asymmetry without the need for electron polarization.  The resulting sensitivity to $a_\tau$ that remains after the loop suppression is derived in
Sec.~\ref{sec:atau_loop}, before we conclude in Sec.~\ref{sec:con}.

\section{Formalism}
\label{sec:formalism}

The interaction of a lepton $\ell$ with a vector or an axial-vector current $j^\mu$ can be parameterized according to~\cite{Jegerlehner:2017gek}
\begin{align}
\langle \ell(p')|j^\mu|\ell(p)\rangle &=  \,e \,\bar{u}(p')\, \Big[\gamma^\mu F_1 + (i F_2 + F_3 \gamma_5)\, \frac{\sigma^{\mu\nu}q_\nu}{2m_\ell}\nonumber \\
& \qquad +
\gamma^\mu\gamma_5 F_4+\frac{q^\mu}{2m_\ell}\big(F_5+\gamma_5 F_6\big)
 \Big]\, u(p)\,,
\end{align}
where the form factors $F_i(s)$, $s=q^2$, depend on the momentum transferred by the photon, $q$, and Gordon identities have been used to reduce the number of form factors as far as possible. The form factor $F_1(s)$ describes the vectorial component of the electromagnetic current $j^\mu=j^\mu_\text{em}$, in which case $F_1(0)=1$ by charge conservation. Likewise, the $s\to 0$ limits of the dipole form factors $F_{2,3}(s)$
 can be directly related to the anomalous magnetic  and electric dipole  moments $a_\ell$ and $d_\ell$, respectively:
\begin{equation}
a_\ell = \Re F_2(0) \,, \qquad d_\ell = \frac{e}{2m_\ell} \Re F_3(0)\,.
\end{equation}
In the case of a conserved current such as $j^\mu_\text{em}$ or the vectorial component of the $Z$-boson current one has
\begin{equation}
 F_5(s)=0\,,\qquad
 F_6(s)=-\frac{4m_\ell^2}{s}F_4(s)\,,
\end{equation}
while, e.g., for the axial-vector component of the $Z$ boson the two additional form factors $F_{5,6}(s)$ are required. In the following, we assume the form of a conserved current, adding the $Z$-boson contributions as separate terms.  

Generalizing the corresponding expressions from Ref.~\cite{Gogniat:2025eom}, we write the cross section for $e^+e^-\to\tau^+\tau^-$ in the following form 
\begin{align}
\label{cross_section_QED}
    \frac{d\sigma}{d\Omega} &= \frac{\alpha_\text{em}^2 \beta_\tau}{4 s \beta_e} \bigg[ |F_1|^2\left(2-\beta_\tau^2 + \beta_e^2\beta_\tau^2\cos^2\theta\right) + |F_2|^2 + 4\Re(F_1F_2^*) \notag\\
    &+ (|F_2|^2+|F_3|^2)\gamma_\tau^2\Big(1-\beta_e^2\beta_\tau^2\cos^2\theta\Big) 
     + \frac{|F_1+F_2|^2}{\gamma_e^2} - |F_3|^2\notag\\
 &+\beta_\tau^2\Big(2-\beta_e^2\sin^2\theta\Big) |F_4|^2   
     \bigg]+\frac{d\sigma_Z}{d\Omega}\,,
\end{align}
with $\beta_\ell = \sqrt{1-4m_\ell^2/s}$, $\gamma_\ell = \sqrt{s}/(2m_\ell)$, and scattering angle $\theta$. The $Z$-boson contributions are given explicitly in \ref{app:Z}.
Similarly, we generalize the spin-dependent contributions 
\begin{align}
\label{sigma_S}
    \frac{d\sigma^S}{d\Omega} &= \frac{\alpha_\text{em}^{2}  \beta_\tau}{8 s \beta_e} \sum_{\pm}\bigg[(s_- \pm s_+)_x \,\big(X_\pm+X_\pm^Z\big) \notag\\
    &+ (s_- \pm s_+)_y \, \big(Y_\pm+Y_\pm^Z\big) 
    + (s_- \pm s_+)_z \, \big(Z_\pm+Z_\pm^Z\big)\bigg]\,,\notag\\
X_+&=\frac{\beta_\tau}{\gamma_\tau}\beta_e^2 \sin\theta \cos\theta \left[ \Re (F_4 F_1^*) + \gamma_\tau^2\Re (F_4 F_2^*) \right]\,,\notag\\
    X_- &= \beta_\tau\beta_e^2 \gamma_\tau \sin\theta \cos\theta \left[ \Im (F_3 F_1^*) + \Im (F_3 F_2^*) \right]\,, \notag\\ 
    Y_+ &= \beta_\tau^{2}\beta_e^2 \gamma_\tau \sin\theta \cos\theta \, \Im (F_2 F_1^*)\,, \nonumber\\
 Y_- &= -\beta_\tau^{2}\beta_e^2 \gamma_\tau  \sin\theta \cos\theta \, \Re (F_4 F_3^*)\,, \nonumber\\
 Z_+ &= \beta_\tau \big(2-\beta_e^2\sin^2\theta\big) \left[ \Re (F_4 F_1^*) + \Re (F_4 F_2^*) \right]\,,\notag\\
    Z_- &= -\beta_\tau \big(1-\beta_e^2\cos^2\theta\big) \left[ \Im (F_3 F_1^*) + \gamma_\tau^{2} \, \Im (F_3 F_2^*) \right]\,,
\end{align}
and 
\begin{align}
\label{sigmaSlambda}
    \frac{d\sigma^{S\lambda}}{d\Omega} &= \frac{d\sigma_Z^{S\lambda}}{d\Omega}+\frac{\alpha_\text{em}^2 \lambda \beta_\tau}{16 s \beta_e} \bigg\{2\beta_\tau\cos\theta\left[ \Re (F_4 F_1^*) + \Re (F_4 F_2^*) \right]\notag\\
    &+\sum_\pm\bigg[ 
                        (s_- \pm s_+)_{x} \, \big(X_\pm^\lambda+X_\pm^{\lambda,Z}\big) \notag\\
                        &+ (s_- \pm s_+)_{y} \, \big(Y_\pm^\lambda+Y_\pm^{\lambda,Z}\big) + (s_- \pm s_+)_{z} \, \big(Z_\pm^\lambda+Z_\pm^{\lambda,Z}\big) 
                    \bigg]\bigg\}\,,\notag\\
     X_+^\lambda &= \frac{\sin \theta}{\gamma_\tau} \left[ |F_1|^2 + (1 + \gamma_\tau^{2}) \, \Re (F_2 F_1^*) + \gamma_\tau^{2} \, |F_2|^2 \right]\,,\notag\\
     X_-^\lambda&=-\beta_\tau^2\gamma_\tau\sin\theta\, \Im (F_4 F_3^*)\,,\notag\\
     Y_+^\lambda &=-\frac{\beta_\tau}{\gamma_\tau}\sin\theta\left[ \Im (F_4 F_1^*) +\gamma_\tau^2 \,\Im (F_4 F_2^*) \right]\,,\notag\\
     Y_-^\lambda &= -\beta_\tau \gamma_\tau \sin \theta \left[ \Re (F_3 F_1^*) + \Re (F_3 F_2^*) \right]\,,\notag\\
     Z_+^\lambda &= \cos \theta \, \Big(|F_1 + F_2|^2+\beta_\tau^2|F_4|^2\Big)\,, \qquad
  Z_-^\lambda=0\,,
\end{align}
where $s_\pm$ denote the spins of $\tau^\pm$ and $\lambda$ the electron polarization. Throughout, we follow the normalization conventions of Ref.~\cite{Gogniat:2025eom}, and the $Z$-boson contributions are provided in \ref{app:Z}.

To isolate the different form factors, and thereby constrain potential BSM contributions 
\begin{equation}
  F_i(s)=F_i^\text{SM}(s)+F_i^\text{BSM}(s)\,,  
\end{equation}
one compelling strategy concerns the measurement of suitably defined asymmetries~\cite{Bernabeu:2004ww,Bernabeu:2006wf,Bernabeu:2007rr,Bernabeu:2008ii}, see \ref{app:asymmetries} for their precise definitions. Among possible asymmetries 
are the normal asymmetry $A_N^\pm$, the longitudinal and transverse asymmetries $A_L^\pm$ and $A_T^\pm$, as well as their $CP$-odd counterparts, $A_{N,F_3}^{\pm}$, $A_{L,F_3}^{\pm}$, and $A_{T,F_3}^{\pm}$.\footnote{The asymmetries as defined in \ref{app:asymmetries} do not have definite $CP$ properties yet. To this end, (anti-)symmetrization over $h=\pm$ is required, see Eq.~\eqref{eq:asymmetries2}.} While measuring $A_N^\pm$, $A_{T,F_3}^{\pm}$, and $A_{L,F_3}^{\pm}$ only requires accessing information on the spin and momenta of the decay products of a final-state $\tau^\pm$, measuring the other asymmetries additionally requires polarized electron beams, which would become available with the polarization upgrade of the SuperKEKB collider~\cite{USBelleIIGroup:2022qro,Aihara:2024zds}.
Using the decomposition in Eqs.~\eqref{sigma_S} and~\eqref{sigmaSlambda}, one obtains
\begin{align}
\label{eq:asymmetries}
A^\pm_T &= \mp \alpha_\pm\frac{\pi^2 \alpha^2_\text{em} \beta_\tau\gamma_\tau }{4 s \beta_e\sigma_\text{tot}} \bigg(\frac{1}{\gamma_\tau^2}|F_1|^2+(2-\beta_\tau^2)\Re (F_2 F_1^*)+ |F_2|^2\notag\\
& \qquad\pm \beta_\tau^2\Im (F_4 F_3^*)\bigg)\,, \nonumber \\
A_{L}^\pm&=\mp \alpha_\pm\frac{\pi \alpha^2_\text{em} \beta_\tau}{2 s \beta_e\sigma_\text{tot}} \bigg(|F_1+F_2|^2+\beta_\tau^2 |F_4|^2\bigg)\,, \nonumber \\
A_{N,F_3}^\pm  &= \alpha_\pm\frac{\pi^2 \alpha^2_\text{em} \beta_\tau^2 \gamma_\tau }{4 s \beta_e\sigma_\text{tot}}\bigg( \Re (F_3 F_1^*+F_3 F_2^*)  \notag \\
&\qquad \mp \Im (F_4 F_2^*) \mp \frac{1}{\gamma_\tau^2} \Im (F_4 F_1^*) \bigg)\,, \nonumber \\
A_N^\pm  &= \pm\alpha_\pm\frac{\pi \alpha_\text{em}^2 \beta_e \beta_\tau^3 \gamma_\tau }{3 s\sigma_\text{tot}}\left(\Im (F_2 F_1^*)\pm \Re (F_4 F_3^*) \right) \,,\nonumber \\
A_{T,F_3}^\pm  &= \alpha_\pm\frac{\pi \alpha^2_\text{em} \beta_e \beta_\tau^2 \gamma_\tau }{3 s\sigma_\text{tot}} \bigg( \Im (F_3 F_1^*+F_3 F_2^*) \,\nonumber \\ 
&\qquad \mp  \Re (F_4 F_2^*)  \mp \frac{1}{\gamma_\tau^2}\Re (F_4 F_1^*) \bigg)\,,\nonumber \\
A_{L,F_3}^\pm &= -\alpha_\pm \frac{\pi \alpha_\text{em}^2 \beta_\tau^2 (3-\beta_e^2)}{3 s \beta_e  \sigma_\text{tot}} \bigg(\Im (F_3 F_1^*) + \gamma_\tau^2\, \Im (F_3 F_2^*) \notag\\
&\qquad\pm 2 \Re (F_4 F_1^* + F_4 F_2^*)  \bigg)\,,
\end{align}
while the $Z$-boson contributions will be given in Sec.~\ref{sec:Zboson}.
In particular, combining $A_L^\pm$ and $A_T^\pm$, one can isolate the real part of the $F_1$--$F_2$ interference
\begin{align}
\Delta A_{TL}^\pm\equiv A_{T}^\pm - \frac{\pi}{2 \gamma_\tau}A^\pm_L  &=\mp \alpha_\pm\frac{\pi^2 \alpha^2_\text{em} \beta_\tau^3 \gamma_\tau }{4 s \beta_e\sigma_\text{tot}} \bigg(\Re (F_2 F_1^*)+ |F_2|^2\notag\\
& \qquad\pm \Im (F_4 F_3^*) -\frac{1}{\gamma_\tau^2} |F_4|^2 \bigg) \,.
\end{align}

In addition, if the decay products of the $\tau^\pm$ are charge-conjugate states, $\alpha_+ = \alpha_- = \alpha_h$, one can further isolate certain terms by performing appropriate (anti-)symmetrizations
\begin{align}
\label{eq:asymmetries2}
\Delta A^h_{TL}&\equiv \frac{\Delta A_{TL}^--\Delta A_{TL}^+}{2} \\
&= \alpha_h\frac{\pi^2 \alpha^2_\text{em} \beta_\tau^3 \gamma_\tau }{4 s \beta_e\sigma_\text{tot}} \bigg(\Re (F_2 F_1^*)+ |F_2|^2
  -\frac{1}{\gamma_\tau^2} |F_4|^2 \bigg) \,, \nonumber \\
A_{N,F_3}^h&\equiv \frac{A_{N,F_3}^+ + A_{N,F_3}^-}{2}  = \alpha_h\frac{\pi^2 \alpha^2_\text{em} \beta_\tau^2 \gamma_\tau }{4 s \beta_e\sigma_\text{tot}} \Re (F_3 F_1^*+F_3 F_2^*)\,,  \nonumber \\
A_N^h&\equiv \frac{A_N^+-A_N^-}{2}  = \alpha_h\frac{\pi \alpha_\text{em}^2 \beta_e \beta_\tau^3 \gamma_\tau }{3 s\sigma_\text{tot}} \Im (F_2 F_1^*)  \,,\nonumber \\
A_{T,F_3}^h &\equiv\frac{A_{T,F_3}^+ +A_{T,F_3}^-}{2}  = \alpha_h \frac{\pi \alpha^2_\text{em} \beta_e \beta_\tau^2 \gamma_\tau }{3 s\sigma_\text{tot}}  \Im (F_3 F_1^*+F_3 F_2^*)\,, \nonumber\\
A_{L,F_3}^h &\equiv \frac{A_{L,F_3}^++A_{L,F_3}^-}{2}\notag \\
&=-\alpha_h \frac{\pi \alpha_\text{em}^2 \beta_\tau^2 (3-\beta_e^2)}{3 s \beta_e \sigma_\text{tot}} \Big(\Im (F_3 F_1^*) + \gamma_\tau^2\, \Im (F_3 F_2^*)\Big)\,,\notag
\end{align}
in particular, the asymmetries $A_{N,F_3}^h$, $A_{T,F_3}^h$, $A_{L,F_3}^h$ are now $CP$ odd, the others $CP$ even. 

Corrections to Eq.~\eqref{eq:asymmetries} arise, e.g., from QED box diagrams, and in case cuts are imposed on the angular measurements~\cite{Gogniat:2025eom}. In addition, subleading effects are expected from the exchange of a virtual $Z$ boson or four-fermion operators parameterizing general BSM effects. It is the purpose of this Letter to quantify the impact of such interactions on the asymmetries of interest.

\section{\texorpdfstring{$Z$}{}-boson contributions}
\label{sec:Zboson}

The impact of neutral weak interactions mediated by a $Z$ boson on the set of asymmetries we are considering consists of two distinct classes of contributions. Most relevant is the $\Order(G_F \alpha_\text{em}, G_F^2)$ tree-level exchange of a $Z$ boson, which we will consider in detail in the following with focus on $\Delta A_{TL}^\pm$.
Relative to the tree-level QED contribution, these corrections scale as 
\begin{equation}
\label{scaling}
\frac{m_e^2 G_F}{4\pi\alpha_\text{em}}\simeq 3\times 10^{-11}\,,\qquad 
\frac{s m_\tau^2 G_F^2}{(4\pi\alpha_\text{em})^2}\simeq 6\times 10^{-6}\,,
\end{equation}
for the interference with tree-level QED and $Z$-boson exchange squared, respectively, where the latter is evaluated for a typical Belle II center-of-mass energy $\sqrt{s_B}=10.58\GeV$. While the chirality suppression with $m_e^2$ renders the interference negligible, the direct $Z$-boson contribution could become relevant when pushing the precision on $a_\tau$ below the $10^{-5}$ level.

A second class of contributions is given by the one-loop exchange of a $Z$ boson impacting the form factors $F_i$ of the electromagnetic vertex. These effects are suppressed by $\alpha_\text{em}/\pi$, thus playing a subleading role in our analysis.
In particular, a contribution that belongs to this class of effects is the one related to the $P$-violating, $T$-even anapole moment $F_4$. $F_4$ is induced in the SM only by a weak loop involving the exchange of a virtual $Z$ boson and results in an $\Order(\alpha_\text{em})$ correction to neutral current exchanges~\cite{Musolf:1990sa}. In particular, since it is possible to remove all the effects sourced by $F_4$ apart from a term proportional to $|F_4|^2$, the only unavoidable impact on $\Delta A_{TL}$ scales as 
\begin{equation}
\frac{1}{\gamma_\tau^2}\bigg(\frac{s G_F}{4\pi^2}\bigg)^2\simeq 10^{-10}\,,
\end{equation}
and can therefore be neglected at Belle II energies.

Accordingly, it is sufficient to concentrate on the effects induced by the exchange of a $Z$ boson at tree level. The full expressions for the cross sections and asymmetries, used for the numerical analysis, are given in \ref{app:Z}, while in the main text we report the leading contributions in the limit $M_Z\to\infty$ (and $m_e\to 0$). 
Partial expressions were computed and provided in Refs.~\cite{Bernabeu:1994wh,Bernabeu:2006wf,Bernabeu:2007rr}.
For the transverse--longitudinal asymmetry we find
\begin{align}
\label{eq:PureZ_contribution}
\Delta A_{TL}^\pm&\simeq \alpha_\pm\frac{G_F g_A^2\beta_\tau^3}{4\gamma_\tau\sigma_\text{tot}}\bigg[ \pi\sqrt{2}\,\alpha_\text{em} \frac{m_e^2}{s}\bigg(1-\frac{\Gamma_Z^2}{M_Z^2}\bigg)
\pm \frac{G_F (g_A^2+g_V^2)s}{8}\bigg]\,,
\end{align}
where we used the conventions
\begin{equation}
g_V=-\frac{1}{2}+2s_W^2\,, \qquad g_A=-\frac{1}{2}\,,
\end{equation}
with weak mixing angle $s_W=\sin\theta_W$.
The scaling~\eqref{scaling} is reproduced once expressing the result in terms of 
\begin{align}
\Re  F_2^\text{eff}(s) &\equiv \mp \frac{4s\beta_e\sigma_\text{tot}}{\pi^2\alpha_\text{em}^2\beta_\tau^3\gamma_\tau \alpha_\pm}\Delta A_{TL}^\pm
\simeq \mp \frac{4(3-\beta_e^2)(3-\beta_\tau^2)}{3\pi\beta_\tau^2\gamma_\tau}\frac{\Delta A_{TL}^\pm}{\alpha_\pm}\notag\\
&\simeq \mp0.68\frac{\Delta A_{TL}^\pm}{\alpha_\pm}
\simeq 2.9\times 10^{-6}\,,
\end{align}
where we used the leading QED contribution from Eq.~\eqref{cross_section_QED} for the total cross section. 
The full result is thus a factor of two smaller than the power counting~\eqref{scaling} would suggest, but still needs to be taken into account for a determination of $a_\tau$ that approaches $10^{-6}$ precision. 

For the other asymmetries we obtain
\begin{align}
    A_{N, F_3}^\pm\big|_Z &= \mp\alpha_\pm\frac{\pi\sqrt{2}\, \alpha_\text{em} G_F g_A g_V \beta_\tau^2}{16\gamma_\tau \sigma_\text{tot}}\frac{\Gamma_Z}{M_Z}\,, \qquad 
    A_N^\pm\big|_Z =  0\,, \nonumber\\
    A_{T, F_3}^\pm\big|_Z  &= \mp\alpha_\pm\frac{\sqrt{2}\,\alpha_\text{em} G_F g_A g_V \beta_\tau^2}{12\gamma_\tau \sigma_\text{tot}}\,,\notag\\
    A_{L, F_3}^\pm\big|_Z  &= \mp\alpha_\pm\frac{\sqrt{2}\,\alpha_\text{em} G_F g_A g_V \beta_\tau^2}{3 \sigma_\text{tot}}\,,
\end{align}
which implies that, after symmetrization, no further contributions from the tree-level exchange of a $Z$ boson remain:
\begin{equation}
A_N^h\big|_Z = A_{N, F_3}^h\big|_Z = A_{T,F_3}^h\big|_Z = A_{L,F_3}^h\big|_Z = 0\,. 
\end{equation}

\section{Tree-level four-fermion operators}
\label{sec:4F_tree}

This section addresses the impact of four-fermion operators on the asymmetries introduced in Sec.~\ref{sec:formalism}. In particular, we will consider the tree-level amplitudes involving LEFT four-fermion operators in the basis from Ref.~\cite{Jenkins:2017jig}:
\begin{align}
\label{eq:4fTL}
\mathcal{L}_{\text{4f}}^\text{LEFT} &\supset C_{prst}^{V,LL} (\bar{e}_{Lp} \gamma^\mu e_{Lr})(\bar{e}_{Ls}\gamma_\mu e_{Lt})+ C_{prst}^{V,RR} (\bar{e}_{Rp} \gamma^\mu e_{Rr})(\bar{e}_{Rs}\gamma_\mu e_{Rt})\nonumber \\
& + C_{prst}^{V,LR} (\bar{e}_{Lp} \gamma^\mu e_{Lr})(\bar{e}_{Rs}\gamma_\mu e_{Rt}) \nonumber \\
& + \big[C_{prst}^{S,RR} (\bar{e}_{Lp}  e_{Rr})(\bar{e}_{Ls}e_{Rt}) + \text{h.c.}\big]\,,
\end{align}
where we dropped a common subscript $ee$ for all Wilson coefficients ($prst$ referring to generation indices) and suppressed the common factor $1/\Lambda^2$ with the BSM scale $\Lambda$.
The operators corresponding to the Wilson coefficients $C_{prst}^{V,LL}$, $C_{prst}^{V,RR}$, and $C_{prst}^{V,LR}$ are Hermitian, which implies that, for each one of them, $C_{prst} = C^*_{rpts}$. In addition, for $LL$ and $RR$ vectorial operators one has $C_{prst}= C_{stpr}$ by construction. As a consequence, it follows that $C_{ee\tau \tau}^{V,XY} = C_{\tau \tau ee}^{V,XY}$ with $X,Y = \{L,R\}$ is a real Wilson coefficient and that $C_{e\tau \tau e}^{V,XX} = C_{\tau e e\tau}^{V,XX}$.

As shown in Ref.~\cite{Jenkins:2017jig}, the vectorial operators include both a $Z$-boson-exchange contribution and local contributions from BSM physics, which can be conveniently parameterized within a SM effective field theory (SMEFT) setup. Similarly, the operators in the class $\mathcal{O}_{prst}^{S,RR}$ can be directly generated by the exchange of a Higgs boson; their effect, however, is expected to be subdominant, as they experience a suppression by powers of $m^2/v^4$ in the SM, or $\Order(m/v/\Lambda^2,v^2/\Lambda^4)$ if the LEFT is embedded in a SMEFT setup, thus rendering them parameterically of the same order as dimension-8 operators in the LEFT expansion parameter $v=1/(\sqrt{2}G_F)^{1/2}\simeq 246\GeV$~\cite{Jenkins:2017jig}.

Starting from the Lagrangian in Eq.~\eqref{eq:4fTL} we compute the contribution to the asymmetries of interest. In analogy to Eq.~\eqref{scaling} we expect a scaling 
\begin{align}
\label{scaling_4f}
    \frac{m_e m_\tau}{4\pi\alpha_\text{em}v^2}\frac{v^2}{\Lambda^2}\Re C&\simeq 2\times 10^{-7} \frac{v^2}{\Lambda^2}\Re C\,,\notag\\
   \frac{G_Fm_\tau^2 s}{(4\pi\alpha_\text{em})^2v^2}\frac{v^2}{\Lambda^2}\Re C&\simeq 8\times 10^{-6} \frac{v^2}{\Lambda^2}\Re C\,,\notag\\
    \frac{m_\tau^2 s}{(4\pi\alpha_\text{em})^2v^4}\frac{v^4}{\Lambda^4}|C|^2&\simeq 1\times 10^{-5} \frac{v^4}{\Lambda^4}|C|^2\,,
\end{align}
for the interference with photon exchange, $Z$-boson exchange, and for quadratic four-fermion contributions, respectively. 
In particular, the interference with tree-level QED is helicity suppressed: the only four-fermion operators that contribute to $\Delta A_{TL}^\pm$ feature opposite helicities, while the QED vertex is helicity conserving, so that an interference term requires an helicity flip for both the $e^+e^-$ and $\tau^+\tau^-$ pair, resulting in the factor $m_e m_\tau$.  

For $\Delta A_{TL}^\pm$, the interference between the LEFT Lagrangian~\eqref{eq:4fTL} and the photon-mediated SM amplitude reads
\begin{align}
\label{4f_gamma}
\Delta A_{TL}^+\big|^{4f}_{\Lambda^2}&= \frac{\alpha_+\pi\alpha_\text{em}\beta_\tau^3}{32\gamma_e\sigma_\text{tot}\Lambda^2}\Re \Big(C^{S,RR}_{ee\tau\tau}+C^{S,RR\,*}_{ee\tau\tau }+C^{V,LR}_{e\tau\tau e}\!+\!C^{V,LR}_{\tau e e\tau }\Big)\,,\notag\\
\Delta A_{TL}^-\big|^{4f}_{\Lambda^2}&= \frac{\alpha_-\pi\alpha_\text{em}\beta_\tau^3}{32\gamma_e\sigma_\text{tot}\Lambda^2}\Re \Big(C^{S,RR}_{ee\tau\tau}+C^{S,RR\,*}_{ee\tau\tau }+C^{V,LR}_{e\tau\tau e}+C^{V,LR}_{\tau e e\tau }\nonumber \\
& \qquad -C^{S,RR}_{ e\tau \tau e}-C^{S,RR\,*}_{\tau ee \tau}\Big)\,,
\end{align}
up to terms that are further suppressed by $m_e$. These leading contributions to $\Delta A_{TL}^\pm$ therefore differ due to those scalar four-fermion operators that consist of two flavor-violating fermionic currents, such as $(\bar{\tau} P_R e)(\bar{e} P_R \tau)$, 
and these contributions are the only ones that remain after antisymmetrization over $h=\pm$:
\begin{equation}
\Re F_2^{\text{eff}}\big|_{\Lambda^2}^{4f}= -\frac{m_e m_\tau}{4\pi \alpha_\text{em}\Lambda^2}\Re \Big(C^{S,RR}_{ e\tau \tau e}+C^{S,RR\,*}_{\tau ee \tau}\Big)\,.
\end{equation}
Their occurrence can be understood from the fact that polarized electron beams can effectively select only a subset of all the possible effects induced by flavor-violating currents. We have checked explicitly  that the $\pm$ asymmetries are once again identical if one considers equally polarized initial-state electrons and positrons, i.e., by setting $|\lambda_+| = |\lambda_-|$. In the relevant case of polarized electrons, however, terms remain that display exactly the scaling anticipated in Eq.~\eqref{scaling_4f}, whose impact goes well beyond the sensitivity that could be achieved in experiment. 

Next, the interference between the SM $Z$-mediated amplitude and effective four-fermion operators leads to
\begin{align}
\label{4f_Z}
\Delta A_{TL}^\pm \big|^{4f}_{M_Z^2\Lambda^2}&= \pm\alpha_\pm \frac{\sqrt{2}\, G_F g_A s\beta_\tau^3}{128\gamma_\tau\sigma_\text{tot}\Lambda^2} \Re\bigg[g_V\Big(2  C_{ee \tau \tau}^{V,RR}+2  C_{e \tau \tau e}^{V,RR} + C_{ee \tau \tau}^{V,LR}\notag\\
&\quad- C_{\tau \tau ee}^{V,LR}-  2  C_{ee \tau \tau}^{V,LL}-2  C_{e \tau \tau e}^{V,LL}\Big) - g_A\Big(2  C_{ee \tau \tau}^{V,RR}+2  C_{e \tau \tau e}^{V,RR}\notag\\&\quad-
C_{ee \tau \tau}^{V,LR} - C_{\tau \tau ee}^{V,LR} + 2  C_{ee \tau \tau}^{V,LL}+2  C_{e \tau \tau e}^{V,LL}\Big)\bigg]\,,
\end{align}
and thus 
\begin{equation}
 \Re F_2^\text{eff}\big|_{M_Z^2 \Lambda^2}^{4f}    
 =\frac{2\sqrt{2}G_F m_\tau^2 s}{(4\pi\alpha_\text{em})^2\Lambda^2}g_A g_{A/V}\Re C
 \simeq 10^{-5}\Re C\frac{v^2}{\Lambda^2}\,.
\end{equation}
Finally, we have $\mathcal{O}(\Lambda^{-4})$ corrections, whose leading terms are
\begin{align}
\label{4f_squared}
\Delta A_{TL}^\pm \big|^{4f}_{\Lambda^4}&= \pm\alpha_\pm\frac{\beta_\tau^3s}{512\gamma_\tau\sigma_\text{tot}\Lambda^4} \\
&\times\bigg[\Big(2\Re C^{V,RR}_{ee\tau\tau}+ 2\Re C^{V,RR}_{e\tau\tau e}- \Re C^{V,LR}_{\tau\tau ee}\Big)^2 \nonumber \\
& \quad + \Big(2\Re C^{V,LL}_{ee\tau\tau}+ 2\Re C^{V,LL}_{e\tau\tau e}- \Re C^{V,LR}_{ee \tau\tau }\Big)^2 \nonumber \\
& \quad \mp \Re C_{e\tau \tau e}^{S,RR } \Big(\Re C_{e\tau \tau e}^{S,RR } + 2\Re C^{V,LR}_{e \tau\tau e} -2 \Re C_{e e\tau \tau}^{S,RR}\Big)  \nonumber \\
& \quad\mp\Re C_{\tau e e \tau}^{S,RR\,*}\Big(\Re C_{\tau e e \tau}^{S,RR\,*} + 2\Re C^{V,LR}_{\tau e e\tau}- 2\Re C_{ e e \tau \tau}^{S,RR\,*}\Big)\nonumber \\
& \quad \mp \Im C_{e\tau \tau e}^{S,RR } \Big(\Im C_{e\tau \tau e}^{S,RR } + 2\Im C^{V,LR}_{e \tau\tau e} -2 \Im C_{e e\tau \tau}^{S,RR}\Big)  \nonumber \\
& \quad \mp\Im C_{\tau e e \tau}^{S,RR\,*}\Big(\Im C_{\tau e e \tau}^{S,RR\,*} + 2\Im C^{V,LR}_{\tau e e\tau}- 2\Im C_{ e e \tau \tau}^{S,RR\,*}\Big)\bigg]\,.\notag
\end{align}
This leaves
\begin{equation}
 \Re F_2^\text{eff} \big|^{4f}_{M_Z^2\Lambda^2}=  \frac{m_\tau^2 s}{32\pi^2\alpha_\text{em}^2\Lambda^4}|C|^2\simeq 0.5\times 10^{-5} \frac{v^4}{\Lambda^4}|C|^2\,, 
\end{equation}
thus confirming the last of the power-counting estimates in Eq.~\eqref{scaling_4f}. In all cases, the contributions from four-fermion operators are severely suppressed, and could only become relevant for Wilson coefficients of $\Order(1)$ and a BSM scale $\Lambda\simeq v$.  These observations strengthen the conclusions reached in Refs.~\cite{Crivellin:2021spu, Gogniat:2025eom}: the leading heavy BSM effects on the asymmetries of interest are induced by dipole operators, with subdominant corrections from four-fermion operators.

In Eqs.~\eqref{4f_gamma}, \eqref{4f_Z}, and \eqref{4f_squared} the $Z$-boson matching contributions to the LEFT Wilson coefficients are understood to be removed to avoid double counting. However, we can use the known matching contributions~\cite{Jenkins:2017jig}
\begin{align}
C^{V,LL}_{ee\tau\tau}&=C^{V,LL}_{e\tau\tau e}=-\sqrt{2}\,G_FZ_L^2\,,\qquad C^{V,RR}_{ee\tau\tau}=C^{V,RR}_{e\tau\tau e}=-\sqrt{2}\, G_FZ_R^2\,,\notag\\
C^{V,LR}_{ee\tau\tau}&=C^{V,LR}_{\tau\tau ee}=-4\sqrt{2}\, G_F Z_L Z_R\,,\qquad C=0\quad \text{otherwise}\,,\notag\\
Z_L&=\frac{g_V+g_A}{2}\,,\qquad Z_R=\frac{g_V-g_A}{2}\,,
\end{align}
as a cross check on our results. Inserting these expressions into Eqs.~\eqref{4f_Z} and~\eqref{4f_squared} reproduces the $\Order(G_F^2)$ terms in Eq.~\eqref{eq:PureZ_contribution}---in the case of Eq.~\eqref{4f_Z} multiplied by a factor two due to the interference. Similarly, one can verify the $\Order(m_e^2)$ term in Eq.~\eqref{eq:PureZ_contribution}.

For the other asymmetries, even fewer relevant effects from four-fermion operators arise. 
 Regarding the normal asymmetry, we find
\begin{equation}
    A_N^\pm\big|^{4f}_{\Lambda^2} = A_N^\pm\big|^{4f}_{M_Z^2\Lambda^2} = A_N^\pm\big|^{4f}_{\Lambda^4} = 0\,,
\end{equation}
whereas for the $CP$-odd normal asymmetry we only find helicity-suppressed contributions, i.e.,
\begin{equation}
    A_{N,F_3}^\pm\big|^{4f}_{\Lambda^2} = A_{N,F_3}^\pm\big|^{4f}_{M_Z^2\Lambda^2} = A_{N,F_3}^\pm\big|^{4f}_{\Lambda^4} = 0\,,
\end{equation}
up to $\Order(m_e)$ corrections. For the longitudinal and transverse $CP$-odd asymmetries the leading contributions cancel after symmetrization; for completeness, we give the expressions prior to symmetrization in~\ref{app:CPodd}.

\section{Loop-level four-fermion operators}
\label{sec:4F_loop}

\begin{figure}[t]
\centering
	\includegraphics[width=0.7\linewidth]{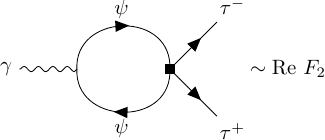}

\caption{Feynman diagrams involving four-fermion operators that contribute to $F_2$. Square dots represent an insertion of the SMEFT dipole operator.}
\label{fig:F2_diagrams}
\end{figure}

In the previous section we discussed the impact of four-fermion operators on the asymmetries of interest at tree level. We found that it leads to relatively weak bounds on four-fermion operators via the transverse--longitudinal asymmetry $\Delta A_{TL}^\pm$, while no effect is generated for the normal asymmetry $A_N^\pm$.
At the loop level, the situation differs rather substantially as far as $A_N^\pm$ is concerned.
The reason behind this statement is simple: closing a four-fermion vertex and attaching it to a photonic line can generate a contribution to $a_\tau$, see Fig.~\ref{fig:F2_diagrams}. If the energy carried by the photon is larger than twice the mass of the virtual fermion pair, $q^2 > 4 m_\psi^2$, the loop develops an imaginary part, which can then be detected by the normal asymmetry $A_N^\pm$.
This represents an interesting opportunity to probe four-fermion operators that are otherwise difficult to access, such as the $4\tau$ operator, or other $\bar{q}q \bar{\tau} \tau$ effective vertices.
In particular, we obtain
\begin{align}
\label{Im_4f}
\Im F_2|_{4f} &= \sum_{\ell = e,\mu,\tau}C_{\ell\tau\tau \ell}^{S,RR} \frac{m_\ell m_\tau}{8\pi \Lambda^2} \beta_\ell 
+ \sum_{q=u,d,s,c,b}C_{q\tau\tau q}^{T,RR} \frac{m_q m_\tau}{\pi \Lambda^2} N_c Q_q\beta_q\,,
\end{align}
where $N_c$ denotes the number of colors, $Q_q$ the quark charges, and $\beta_q$ their phase-space factors. $C_{q\tau \tau q}^{T,RR}$ is the Wilson coefficient associated to the LEFT operators $\mathcal{O}_{eu,prst}^{T,RR} = (\bar{e}_L \sigma^{\mu\nu} e_R)(\bar{u}_L \sigma_{\mu\nu} u_R)$ and $\mathcal{O}_{ed,prst}^{T,RR} = (\bar{e}_L \sigma^{\mu\nu} e_R)(\bar{d}_L \sigma_{\mu\nu} d_R)$ for the choice of the flavor indices given by $prst=q\tau \tau q$.  

Assuming a measurement of $\Im F_2$ with a precision of $10^{-6}$, Eq.~\eqref{Im_4f} then yields the following bounds:
\begin{align}
\Lambda/\sqrt{C_{\ell\tau\tau\ell}^{S,RR}} &> \{343 , 86 ,6  \}  \GeV &\text{for}&& \ell &=\{\tau, \mu ,e\}\, ,\nonumber \\
\Lambda/\sqrt{C_{q\tau\tau q}^{T,RR}} &> \{1181 , 51 \}\GeV &\text{for} &&q &=\{c,u\} \,,\notag\\
\Lambda/\sqrt{C_{q \tau\tau q}^{T,RR}} &> \{1201  , 231 , 52  \} \GeV&\text{for}&& q &=\{b, s ,d\}\,,
\end{align}
where of course a derivation of rigorous bounds for the light quarks $q=u,d,s$ would require the consideration of nonperturbative effects. 
The bound on $C_{\tau\tau\tau\tau}^{S,RR}$ is, to the best of our knowledge, novel, while the bounds on the second-generation up-type quarks are of the same order of magnitude as the results existing in the literature~\cite{Hiller:2025hpf}.

Regarding the transverse--longitudinal asymmetry, at the one loop-level one should consider two effects. The first one consists of loop corrections to the four-fermion operators. With the expected sensitivity, these effects are negligible, as they simply result in a further loop suppression of the already marginal effects discussed in Sec.~\ref{sec:4F_tree}.
The second class of effects arises from generating a contribution to $\Re F_2$ at the loop level. The leading contributions to such a quantity are those induced by the renormalization group of the corresponding LEFT Wilson coefficients~\cite{Jenkins:2017dyc}.
We find:
\begin{align}
\Re F_2|_{4f} &= \sum_{\ell} \frac{m_\ell m_\tau}{8\pi^2 \Lambda^2} \beta_\ell \bigg[-2C_{\ell\tau\tau \ell}^{V,LR}\\ 
&\quad + C_{\ell\tau\tau \ell}^{S,RR}\bigg(2+ \log \frac{\Lambda^2}{m_\ell^2}-\beta_\ell \log\frac{\beta_\ell}{1-\beta_\ell}\bigg)\bigg]\nonumber \\
&+\sum_{q}C_{q\tau\tau q}^{T,RR} \frac{m_q m_\tau}{\pi^2 \Lambda^2} N_c Q_q \bigg(2+ \log \frac{\Lambda^2}{m_q^2}-\beta_q \log\frac{\beta_q}{1-\beta_q}\bigg)\,,\notag
\end{align}
resulting in the following bounds (under the assumption of a measurement of $\Re F_2$ at the level of $10^{-6}$)
\begin{align}
\Lambda/\sqrt{C_{\ell\tau\tau\ell}^{S,RR}} &> \{645 , 129 ,2.8  \}  \GeV&\text{for}&& \ell &=\{\tau, \mu ,e\}\, ,\nonumber \\
\Lambda/\sqrt{C_{\ell\tau\tau\ell}^{V,LR}} &> \{281 , 69 ,5  \}  \GeV&\text{for}&& \ell &=\{\tau, \mu ,e\}\, ,\nonumber \\
\Lambda/\sqrt{C_{q\tau\tau q}^{T,RR}} &> \{975, 83 \}\GeV &\text{for}&& q &=\{c,u\} \,,\notag\\
\Lambda/\sqrt{C_{q \tau\tau q}^{T,RR}} &> \{3292,397,71\} \GeV&\text{for}&& q &=\{b, s ,d\}\,.
\end{align}
Again, some semileptonic $\bar{q}q\bar{\tau}\tau$ Wilson coefficients could be constrained at the same order of magnitude as in high-energy collisions, while constraints on $4\tau$ operators would be novel.

\section{Probing \texorpdfstring{$a_\tau$}{} beyond the Schwinger term via \texorpdfstring{$A_N^\pm$}{}}
\label{sec:atau_loop}

\begin{figure}[t]
\centering
	\includegraphics[width=0.9\linewidth]{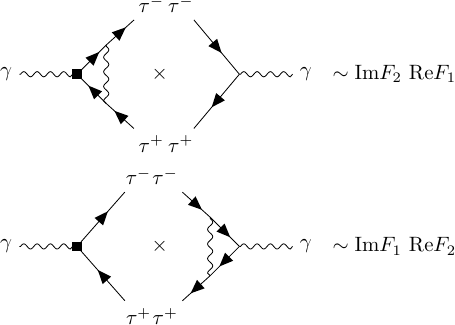}

\caption{Feynman diagrams contributing to $\Im  (F_2 F_1^*)$. Square dots represent an insertion of the SMEFT dipole operator. Other topologies differing for the placement of the SMEFT dipole operator in the loop are not displayed, but were consistently included in the computation.}
\label{fig:AN_diagrams}
\end{figure}

Within SMEFT, the leading contribution to the magnetic dipole moment of the $\tau$ lepton is induced by the local operators
\begin{align}
\label{eq:L_Dip}
\mathcal{L} &\supset
\bar\ell_L \sigma^{\mu\nu}e_{R}
\left(
\frac{C^\ell_{eB}}{\Lambda^2}
H B_{\mu\nu} +
\frac{C^\ell_{eW}}{\Lambda^2}
\tau^I H W_{\mu\nu}^I
\right)
+ {\text{h.c.}}\,,
\end{align}
yielding, at tree level,
\begin{align}
\Delta a_\tau  &\simeq \frac{4m_\tau v}{e\sqrt{2}\Lambda^2} \,
\Re C^\tau_{e\gamma}\,,\qquad C_{e\gamma}=c_W C_{eB} - s_W C_{eW}\,.
\label{eq:Delta_a_ell}
\end{align}
This quantity, being generated by local operators, is necessarily real and can be constrained by measuring the asymmetry $\Delta A_{TL}^\pm$. 
However, in analogy to the loop-level four-fermion insertions discussed in Sec.~\ref{sec:4F_loop}, also the dipole operator generates an imaginary part at one-loop order, which allows for a measurement via the normal asymmetry, at the expense of an additional suppression by a loop factor. 
The relevant quantity sensitive to the SMEFT dipole operator
 is  $\Im (F_2 F_1^*) = \Im  F_2 \Re F_1-\Im  F_1 \Re F_2$, which is, indeed, generated by the interference of the imaginary part of a loop-level diagram involving the SMEFT dipole with the tree-level $F_1$ in the SM, and by the interference of the imaginary part of the SM contribution to $F_1$ and the tree-level SMEFT contribution to $F_2$, see Fig.~\ref{fig:AN_diagrams}.

Considering both effect together, we find the following contribution to $\Im  (F_2 F_1^*)$:
\begin{align}
\Im (F_2 F_1^*) &= - \alpha_\text{em}\frac{5-3/\gamma_\tau^2}{4\beta_\tau}\Delta a_\tau 
\simeq -0.009\,\Delta a_\tau\,.
\end{align}
Including, in addition, the conversion from $A_N^\pm$ to $\Im F_2^\text{eff}$ at Belle-II energies,  it follows that testing $a_\tau$ at the level of the Schwinger term would require a measurement of $A_N^\pm$ with a precision of $0.7\times 10^{-5}$.

\section{Conclusions and outlook}
\label{sec:con}

In this Letter we assessed  the impact of tree-level $Z$-boson exchange and four-fermion operators on asymmetries in $e^+e^-\to\tau^+\tau^-$ constructed to project out the dipole moments of the $\tau$ lepton. For the $Z$-boson contribution, the interference with tree-level QED is helicity suppressed, while the quadratic contribution can affect the determination of $a_\tau$ at the level of $3\times 10^{-6}$. Other asymmetries, sensitive to $d_\tau$, remain unaffected after symmetrization. 
For four-fermion operators, we again find a helicity suppression of the interference with tree-level QED, estimating the largest possible effects as $10^{-5}C v^2/\Lambda^2$. 

Beyond tree level, effective operator insertions also generate imaginary parts, which can be probed via the normal asymmetry without the need for electron polarization. For the dipole operator, this presents an opportunity to access  
$a_\tau$ via its impact at the loop level on $\Im F_2$. If the latter were measured with a precision of at least $10^{-5}$, $a_\tau$ could be tested at the level of the Schwinger term. 

Similarly, also four-fermion operators could be constrained from one-loop diagrams, both via their impact on $\Re F_2$ and $\Im F_2$. The projected precision reflects the additional loop suppression, but in some cases, such as the $4\tau$ operators $C_{\tau\tau\tau\tau}^{S,RR}$ and $C_{\tau\tau\tau\tau}^{V,LR}$, one could still place the best constraints available. Moreover, this
strategy could be repeated, for $\Im F_2$, at future lepton colliders, while for $\Re F_2$ polarized beams would be required. In particular, for energies larger than the di-top threshold, our analysis could be employed in order to test $\tau\tau tt$ or $4t$ four-fermion operators, as motivated  
by UV scenarios that address the flavor puzzle by larger couplings to the third generation of fermions.

\section*{Acknowledgments}

Financial support by the SNSF (Project No.\ TMCG-2\_213690) is gratefully acknowledged. This research was supported by the Munich Institute for Astro-, Particle and BioPhysics (MIAPbP) which is funded by the Deutsche Forschungsgemeinschaft (DFG, German Research Foundation) under Germany's Excellence Strategy -- EXC-2094 -- 390783311.

\appendix

\section{\texorpdfstring{$Z$}{}-boson contributions}
\label{app:Z}

In this appendix we collect the explicit expressions for $Z$-boson contributions (both interference with tree-level QED and quadratic terms):
\begin{align}
 D_Z\frac{d\sigma_Z}{d\Omega}&=\frac{\sqrt{2}\,\alpha_\text{em} G_F \beta_\tau}{8\pi\beta_e}N_Z 
\Big[2 g_A^2 \beta_e \beta_\tau \cos \theta\notag\\
&\quad +g_V^2 \Big(3-\beta_\tau^2 - \beta_e^2+\beta_e^2\beta_\tau^2 \cos^2\theta\Big)\Big] \nonumber\\
&+\frac{G_F^2\beta_\tau}{32\pi^2\beta_e}s M_Z^4\bigg[g_V^4  \Big(3-\beta_\tau^2 - \beta_e^2+ \beta_e^2 \beta_\tau^2 \cos^2 \theta\Big)\notag\\ 
&\quad+g_A^2 g_V^2  \Big(2\beta_\tau^2+2\beta_e^2+ 8 \beta_e \beta_\tau \cos \theta- 2 \beta_e^2 \beta_\tau^2 \sin^2 \theta\Big) \notag \\
& \quad +g_A^4 \bigg(\frac{N_Z^2}{M_Z^8\gamma_e^2\gamma_\tau^2}+\beta_e^2\beta_\tau^2(1+\cos^2\theta)\bigg) \bigg]\,, \nonumber \\
D_Z X_+^Z&=- \frac{\beta_e}{\gamma_\tau}\sin\theta\bigg[ \frac{\sqrt{2}\,G_Fg_A g_V}{4\pi\alpha_\text{em}}s N_Z
 \Big(2 + \beta_e \beta_\tau \cos \theta \Big) \notag\\
&\quad+ \frac{G_F^2 g_A g_V}{8\pi^2\alpha_\text{em}^2}s^2M_Z^4\Big(2g_V^2 +  \big(g_V^2+g_A^2\big) \beta_e \beta_\tau \cos \theta\Big)\bigg]\,,\notag\\
 X_-^Z&=0\,,\notag\\
 D_Z Y_+^Z&= \frac{\sqrt{2}\,G_Fg_A^2}{4\pi\alpha_\text{em}} \frac{sM_Z^3 \Gamma_Z}{\gamma_\tau}\beta_e \beta_\tau \sin \theta\,,\qquad
 Y_-^Z= 0\,,\notag\\
 D_Z Z_+^Z&=-\frac{\sqrt{2}\,G_Fg_A g_V}{4\pi\alpha_\text{em}}s N_Z \Big[ (2- \beta_e^2\sin^2 \theta) \beta_\tau+2  \beta_e  \cos \theta\Big] \notag\\
 &- \frac{G_F^2g_A g_V}{16\pi^2\alpha_\text{em}^2}s^2M_Z^4\bigg[4  \beta_e \big(g_V^2+g_A^2\beta_\tau^2\big)\cos \theta  \notag\\
 &\quad+\beta_\tau\Big(3 g_A^2 \beta_e^2 - g_V^2(\beta_e^2-4)+ (g_V^2+g_A^2)\beta_e^2\cos 2\theta\Big)\bigg]\,,\notag\\
 Z_-^Z&=0 \,,\notag\\
 D_Z \frac{d\sigma_Z^{S\lambda}}{d\Omega}&= -\frac{\sqrt{2}\,\alpha_\text{em} G_F \beta_\tau}{64\pi }N_Z g_A g_V\Big[ \frac{2 \beta_\tau}{\beta_e} \cos \theta + 2-\beta_\tau^2\sin^2 \theta \Big]\nonumber \\
 &+\frac{G_F^2 \beta_\tau}{128\pi^2\beta_e \gamma_\tau^2} s M_Z^4 g_A g_V \nonumber\\
 & \quad \times\bigg[g_A^2 \Big(1-\gamma_\tau^2 (1+ 2 \beta_e \beta_\tau \cos \theta + \beta_\tau^2\cos^2 \theta)\Big) \nonumber \\
 & \quad-g_V^2 \Big(\beta_e + \gamma_\tau^2(\beta_e + 2 \beta_\tau \cos \theta + \beta_e \beta_\tau^2 \cos^2 \theta)\Big) \bigg]\,,\notag\\
  D_ZX_+^{\lambda, Z}&=\frac{\sin \theta}{\gamma_\tau}\bigg[\frac{\sqrt{2}\,G_F}{4\pi\alpha_\text{em}}s N_Z \Big(2 g_V^2 + g_A^2 \beta_e \beta_\tau \cos \theta \Big)\notag\\
  &\quad+\frac{G_F^2g_V^2}{8\pi^2\alpha_\text{em}^2}
  s^2M_Z^4\Big(2 g_A^2  \beta_e \beta_\tau \cos \theta+   \big(g_V^2+g_A^2\beta_e^2\big)\Big)\bigg] \,,\nonumber\\
 D_ZX_-^{\lambda, Z}&= \frac{\sin \theta}{\gamma_\tau \gamma_e^2} \bigg[\frac{\sqrt{2}\,G_Fg_A^2}{4\pi\alpha_\text{em}}N_Z^2\frac{s}{M_Z^4} +\frac{G_F^2 g_A^2 g_V^2}{8\pi^2\alpha_\text{em}^2}s^2N_Z\bigg]\,,\notag\\
 D_ZY_+^{\lambda, Z}&= - \frac{\sqrt{2}\,G_F g_V g_A}{8\pi\alpha_\text{em}} sM_Z^3\Gamma_Z\frac{\beta_\tau}{\gamma_\tau}  \sin \theta\,,\qquad 
 Y_-^{\lambda, Z}= 0\,,\notag\\
 D_ZZ_+^{\lambda, Z}&= \frac{\sqrt{2}\,G_F}{4\pi\alpha_\text{em}}s N_Z \Big[g_A^2(1+\cos^2 \theta)\beta_e \beta_\tau  + 2 g_V^2 \cos \theta\Big]\notag\\
 &+\frac{G_F^2}{8\pi^2\alpha_\text{em}^2}  s^2M_Z^4\Big[2g_A^2 g_V^2(1+\cos^2 \theta)\beta_e \beta_\tau \notag\\
 & \quad +(g_V^2 + g_A^2 \beta_e^2)(g_V^2 + g_A^2 \beta_\tau^2)\cos \theta\Big]\,,\notag\\
 D_ZZ_-^{\lambda, Z}&= \frac{\cos \theta}{\gamma_e^2 \gamma_\tau^2} \bigg[ \frac{\sqrt{2}\,G_Fg_A^2}{4\pi\alpha_\text{em}}N_Z^2\frac{s}{M_Z^4}  
 +\frac{G_F^2g_A^2 g_V^2}{8\pi^2\alpha_\text{em}^2}s^2N_Z \bigg]\,,
\end{align}
where 
\begin{equation}
    D_Z=\big(s-M_Z^2\big)^2+\Gamma_Z^2M_Z^2\,,\qquad N_Z=M_Z^2(s-M_Z^2)\,.
\end{equation}

\section{Asymmetries}
\label{app:asymmetries}

Using the notation of Ref.~\cite{Gogniat:2025eom}, we define the spin components of $\tau^\pm$ via
\begin{equation}
    \mathbf{s}_\pm = \mp \alpha_\pm (\sin \theta^*_\pm \cos \phi_\pm, \sin\theta^*_\pm \sin\phi_\pm, \cos\theta^*_\pm)\,,
\end{equation}
where the polarization analyzer $\alpha_\pm$ depends on the semileptonic decay channel $\tau^\pm\to h^\pm\nu_\tau$ into a hadron $h^\pm$. First, the normal asymmetry 
\begin{equation}
     A_N^\pm =\frac{1}{\sigma_\text{tot}}\bigg[\int_{\pi}^{2\pi} d\phi_{\pm} \frac{d\sigma^S_{\text{FB}}}{d\phi_{\pm}}
     -\int_{0}^{\pi} d\phi_{\pm} \frac{d\sigma^S_{\text{FB}}}{d\phi_{\pm}}\bigg] 
     \label{eq:AN}
\end{equation}
is defined in terms of the forward--backward (FB) asymmetry 
\begin{equation}
\label{sigma_FB}
    d\sigma_\text{FB}^S = \bigg[\int_{0}^{1} dz \frac{d\sigma^S}{d\Omega} - \int_{-1}^{0} dz \frac{d\sigma^S}{d\Omega}\bigg]\,,
\end{equation}
with \( z = \cos \theta\), and integrating over all remaining angles, including a factor $4/(4\pi)^2$~\cite{Bernabeu:2007rr,Tsai:1971vv} that accounts for $\tau$ spins and angular range in the $\tau^\pm\to h^\pm\nu_\tau$ decays. Similarly, we define 
\begin{align}
    A_{T,F_3}^\pm &= \frac{1}{\sigma_{\text{tot}}}\bigg[\int_{-\frac{\pi}{2}}^{\frac{\pi}{2}} d\phi_{\pm} \, \frac{d\sigma_\text{FB}^S}{d\phi_{\pm}}-
    \int_{\frac{\pi}{2}}^{\frac{3\pi}{2}} d\phi_{\pm} \, \frac{d\sigma_\text{FB}^S}{d\phi_{\pm}}\bigg]\,,\notag\\
    A_{L,F_3}^\pm &=\frac{1}{\sigma_{\text{tot}}}\bigg[\int_{0}^{1} dz_\pm^* \, \frac{d\sigma^S}{dz_\pm^*}-
    \int_{-1}^{0} dz_\pm^* \, \frac{d\sigma^S}{dz_\pm^*}\bigg]\,.
\end{align}

For the asymmetries requiring electron polarization we define 
\begin{equation}
    d\sigma_{\text{pol}}^S = d\sigma^{S\lambda}|_{\lambda=1} - d\sigma^{S\lambda}|_{\lambda=-1}\,,
\end{equation}
and
\begin{align}
A_{N, F_3}^\pm &= \frac{1}{\sigma_\text{tot}}\bigg[\int_{\pi}^{2\pi} d\phi_{\pm} \frac{d\sigma_{\text{pol}}^S}{d\phi_{\pm}}- \int_{0}^{\pi} d\phi_{\pm} \frac{d\sigma_{\text{pol}}^S}{d\phi_{\pm}}\bigg]\,,\notag\\
A_{T}^\pm&=\frac{1}{\sigma_\text{tot}}\bigg[\int_{-\frac{\pi}{2}}^{\frac{\pi}{2}} d\phi_{\pm} \frac{d\sigma_{\text{pol}}^S}{d\phi_{\pm}}- \int_{\frac{\pi}{2}}^{\frac{3\pi}{2}} d\phi_{\pm} \frac{d\sigma_{\text{pol}}^S}{d\phi_{\pm}}\bigg]\,,\notag\\
A_{L}^\pm&=\frac{1}{\sigma_{\text{tot}}}\bigg[\int_{0}^{1} dz_\pm^* \, \frac{d\sigma^S_\text{FB,pol}}{dz_\pm^*}-
    \int_{-1}^{0} dz_\pm^* \, \frac{d\sigma^S_\text{FB,pol}}{dz_\pm^*}\bigg]\,,
\end{align}
where
\begin{equation}
\label{sigma_FB_pol}
    d\sigma_\text{FB,pol}^S = \bigg[\int_{0}^{1} dz \frac{d\sigma^S_\text{pol}}{d\Omega} - \int_{-1}^{0} dz \frac{d\sigma^S_\text{pol}}{d\Omega}\bigg]\,.
\end{equation}

\section{$CP$-odd asymmetries}
\label{app:CPodd}

As discussed in Sec.~\ref{sec:4F_tree}, the tree-level four-fermion contributions to $A_{T,F_3}^\pm$, $A_{L,F_3}^\pm$ vanish after symmetrization. The explicit expressions read (at leading order in $M_Z\to\infty$ and for $m_e\to 0$):
\begin{align}
    A_{L,F_3}^\pm\big|^{4f}_{\Lambda^2} &=4\gamma_\tau A_{T,F_3}^\pm\big|^{4f}_{\Lambda^2} = \mp\alpha_\pm\frac{\alpha_\text{em} \beta_\tau^2}{12\sigma_\text{tot}\Lambda^2}\Re\Big(2 C_{ee \tau \tau}^{V,RR}+2 C_{e \tau \tau e}^{V,RR}\notag\\
    &\quad+C_{e e \tau \tau }^{V,LR} -2 C_{ee \tau \tau}^{V,LL}-2C_{e \tau \tau e}^{V,LL}-C_{\tau \tau e e}^{V,LR}\Big)\,, \nonumber\\
   A_{L,F_3}^\pm\big|^{4f}_{M_Z^2\Lambda^2} &= 4\gamma_\tau A_{T,F_3}^\pm\big|^{4f}_{M_Z^2\Lambda^2} = \pm \alpha_\pm\frac{\sqrt{2}\,G_F  \beta_\tau^2 s}{48\pi \sigma_\text{tot}  \Lambda^2}\Re\bigg[g_A^2\Big (2 C_{ee \tau \tau}^{V,RR}\notag\\
   &\quad\quad +2 C_{e \tau \tau e}^{V,RR}-C_{e e \tau \tau }^{V,LR}
   -2 C_{ee \tau \tau}^{V,LL}-2C_{e \tau \tau e}^{V,LL}+C_{\tau \tau e e}^{V,LR}\Big) \nonumber\\
   &\quad -4 g_A g_V \Big( C_{ee \tau \tau}^{V,RR}+ C_{e \tau \tau e}^{V,RR}+ C_{ee \tau \tau}^{V,LL}+C_{e \tau \tau e}^{V,LL}\Big) \nonumber\\
   &\quad +g_V^2 \Big(2 C_{ee \tau \tau}^{V,RR}+2 C_{e \tau \tau e}^{V,RR}+C_{e e \tau \tau }^{V,LR}\notag\\
   &\quad\quad-2 C_{ee \tau \tau}^{V,LL}-2C_{e \tau \tau e}^{V,LL}-C_{\tau \tau e e}^{V,LR}\Big)\bigg]\,, \nonumber\\
   A_{L,F_3}^\pm\big|^{4f}_{\Lambda^4} &=4\gamma_\tau A_{T,F_3}^\pm\big|^{4f}_{\Lambda^4} = \mp \alpha_\pm \frac{\beta_\tau^2 s}{96 \pi\sigma_\text{tot}\Lambda^4}\notag\\
   &\times \bigg[  4 \Big(\Re C_{ee \tau \tau}^{V,RR}+\Re C_{e \tau \tau e}^{V,RR}\Big)^2 +\Big(\Re C_{e e \tau \tau }^{V,LR}\Big)^2 \notag\\
   &\quad 
   - 4
   \Big(\Re C_{ee \tau \tau}^{V,LL}+\Re C_{e \tau \tau e}^{V,LL}\Big)^2-\Big(\Re C_{\tau \tau e e}^{V,LR}\Big)^2 \bigg]\,.
\end{align}

\bibliographystyle{apsrev4-1_mod}
\balance
\biboptions{sort&compress}
\bibliography{tau}

\end{document}